\documentclass[prl,twocolumn,aps,epsfig,prbbib]{revtex4}
\usepackage{graphicx}

\begin{document}

\title{Electronic band structure and  
carrier effective mass in calcium aluminates}

\author{Julia E. Medvedeva}\email[]{E-mail:juliaem@umr.edu}
\author{Emily N. Teasley}
\author{Michael D. Hoffman}

\affiliation{Department of Physics, University of Missouri--Rolla, Rolla, MO 65409}

%\date{\today}

\begin{abstract}
First-principles electronic band structure investigations of five compounds of
the CaO-Al$_2$O$_3$ family, 3CaO$\cdot$Al$_2$O$_3$, 12CaO$\cdot$7Al$_2$O$_3$, 
CaO$\cdot$Al$_2$O$_3$, CaO$\cdot$2Al$_2$O$_3$ and CaO$\cdot$6Al$_2$O$_3$,  
as well as CaO and $\alpha$-, $\theta$- and $\kappa$-Al$_2$O$_3$ are performed. 
We find that the conduction band in the complex oxides is formed from the oxygen antibonding 
$p$-states and, although the band gap in Al$_2$O$_3$ is almost twice larger 
than in CaO, the $s$-states of {\it both} cations.
Such a hybrid nature of the conduction band leads to isotropic 
electron effective masses which are nearly the same for all compounds investigated.
This insensitivity of the effective mass to variations in the composition and structure
suggests that upon a proper degenerate doping, both amorphous and crystalline phases of the materials
will possess mobile extra electrons. 

{\bf PACS number(s)}: 71.20.-b  
\end{abstract}

\maketitle

\subsection{Introduction}

Oxides of the main group metals such as CaO and Al$_2$O$_3$ are known for their superior 
refractory properties and until recently 
these materials seemed to be inappropriate choice to serve as electrical conductors. 
Hence, the discovery \cite{Hosono-original} of an insulator-conductor conversion 
in 12CaO$\cdot$7Al$_2$O$_3$ (12C7A or mayenite), a member of Portland 
cements, generated a lot of excitement and fueled the quest for new directions towards 
inexpensive and environmentally friendly materials for (opto)electronic applications.

The remarkable electrical properties of 12CaO$\cdot$7Al$_2$O$_3$ --
currently, conductivities as high as 1700 S/cm were achieved 
\cite{Bertoni-thesis,Kim} and various insulator-to-metal conversion approaches 
were employed 
\cite{Hosono-original,HosonoAdvanced,Science,myPRL,Bertoni-thesis,Kim,Kim123}
-- originate from its unique structural feature, namely, 
the presence of so-called ``free'' oxygen ions located inside spacious cages 
of $\sim$5.6 \AA \, in diameter. 
Introduction of other charge-balance ions (H$^-$, OH$^-$, Cl$^-$, F$^-$)
into the cages or their 
reduction leads to a wide range of semiconducting to metallic behavior which can be controlled 
via the concentration of these ions.
Accurate band structure calculations have revealed the origin of the observed unusual 
phenomena and also allowed predictions of ways to vary the conductivity over
several orders of magnitude \cite{myPRL,myAPL,MarianaJAP,MarianaSi}.
Furthermore, it was demonstrated \cite{my1epl} that 12CaO$\cdot$7Al$_2$O$_3$ is the first 
of a conceptually 
new class of transparent conductors where the structural peculiarities and the resulting 
electronic band structure features suggest an approach to achieve good electrical conductivity 
without compromising their optical properties -- the bottleneck
in conventional transparent conducting oxides \cite{my1epl,myreview}.

In this work we focus on the structural, electronic and optical properties of the other four 
members of the CaO-Al$_2$O$_3$ family -- 3CaO$\cdot$Al$_2$O$_3$ (3CA), 
CaO$\cdot$Al$_2$O$_3$ (CA), CaO$\cdot$2Al$_2$O$_3$ (C2A or grossite) and CaO$\cdot$6Al$_2$O$_3$ 
(C6A or hibonite) -- and compare with those of 12C7A, CaO and  
$\alpha$--, $\theta$-- and $\kappa$--Al$_2$O$_3$.
Based on the results of the first-principles density functional investigations, 
we find that despite variations in the composition and 
structural diversity in these calcium aluminate compounds, their 
electron effective masses are nearly the same
and are isotropic. The later suggests that isotropic character of 
the electron transport can be achieved upon proper carrier generation 
in these complex multi-cation oxides.
We compare the electronic band structure features of 
calcium aluminates 
with those of the conventional transparent conducting oxides (TCO)
and discuss the advantages of CaO-Al$_2$O$_3$ compounds 
as candidates for novel TCO hosts.

\subsection{Crystal structure}

Calcium aluminates (also known as high alumina cements) have a rich phase diagram 
which includes five lime-alumina compounds -- 3CA, 12C7A, CA, C2A and C6A 
\cite{Rankin,Wisnyj,Barnes,Morozova,Hallstedt,Singh,Goktas,Eriksson,Gulgun,Tas,Yi}. 
Among them, 12C7A has the lowest melting point (1722 $^{\circ}$C) 
and C6A has the highest (2156 $^{\circ}$C) which is comparable to the one 
of pure alumina (2327 $^{\circ}$C). 
Such a superior refractory properties of these materials make them attractive 
for various applications as ceramics and glasses as well as in cement 
and steel industries \cite{Kopanda,Shelby}.
Calcium aluminates have been grown by several techniques
including solid state reactions (sintering), sol-gel technologies and
self-propagating combustion synthesis, see \cite{Kopanda,Tas,Yi} and references therein.

CaO-Al$_2$O$_3$ family has two cubic, two monoclinic and one hexagonal 
structure.
Table \ref{table} lists the lattice space groups of the compounds 
along with the number of formula units in the unit cell
(used in our band structure calculations), the average cation-anion distances 
and oxygen coordination of the cations.
Calcium aluminates exhibit a variety of structural peculiarities.
The above mentioned 12C7A with nanoporous cages has loosely bound oxygen 
O$^{2-}$ ions inside two out of the total 12 cages 
(in the conventional unit cell). 
These ``free'' anions can be easily substituted with F$^-$, Cl$^-$ or OH$^-$ 
\cite{Jeev,Chatterji,Williams,Zhmoidin} or reduced \cite{Science}.
In addition, aliovalent substitution of Al$^{3+}$ with Si$^{4+}$ results in 
an increase of the number of the free oxygen ions and in the formation of
oxygen radicals such as O$^-_2$ and O$^{2-}_2$ inside the cages 
\cite{MarianaSi,maySi}.
Tricalcium aluminate, 3CA, consists of six-fold rings of AlO$_4$ tetrahedra surrounding 
structural voids of 1.5 \AA \, in diameter; Ca ions join these rings together \cite{Mondal}. 
In calcium hexaluminate, C6A, there are double layers of pure AlO$_4$ 
and ``penta-coordinated'' Al$^{3+}$ ions which are displaced from the trigonal 
pyramidal site center \cite{Utsunomiya,Du,Hofmeister}. 
These structural peculiarities may result in specific features in 
the electronic band structure and also
may suggest possible ways for efficient carrier generation. 
For example, empty spaces can serve as sites for dopants or guest atoms.
Atoms from the structurally 
distinct layers or with unusual coordination may facilitate a defect formation
or may be the target for substitutional doping.
Ions that are loosely bound to the host framework can be easily reduced 
to provide extra electrons which balance the charge neutrality and so may
lead to electrical conductivity.

\subsection{Methods}

The electronic band structure calculations of calcium aluminates 
were performed using two density functional methods within
the local density approximation.
First, the linear muffin-tin orbital method (LMTO) in the atomic sphere 
approximation \cite{LMTO} was employed.
For these calculations, lattice parameters and atomic positions were fixed to the experimental 
values (3CA \cite{Mondal}, 12C7A \cite{Christensen}, CA \cite{Horkner}, C2A \cite{Goodwin}, 
C6A \cite{Utsunomiya}, monoclinic $\theta$-Al$_2$O$_3$ \cite{Husson}
and orthorhombic $\kappa$-Al$_2$O$_3$ \cite{Ollivier}). 
The muffin-tin radii are 3.0-4.0 a.u. for Ca, 1.9-2.3 a.u. for Al 
and 1.7-2.2 a.u. for O atoms. Because these structures are not closely packed,  
we included 2, 320, 85, 84, 40, 70, 8, 23 and 44 empty spheres 
to fill the open space in CaO, 3CA, 12C7A, CA, C2A, C6A, $\alpha$-Al$_2$O$_3$, 
$\theta$-Al$_2$O$_3$ and $\kappa$-Al$_2$O$_3$ structure, respectively. 
The number of irreducible {\bf k}-points in the Brillouin zone 
was in the range of 11-170 points. 
 
In addition, we employed the highly-precise full-potential linearized 
augmented plane-wave (FLAPW) method \cite{FLAPW,FLAPW1} 
to calculate accurately the atomic contributions 
to the conduction band wavefunctions in CaO, 12C7A, C2A, 
C6A and Al$_2$O$_3$. 3CA and CA were not included in these studies 
because of the large number of atoms in their unit cells, 264 and 84, 
respectively, which makes the calculations computationally challenging. 
For each structure investigated within FLAPW method, the internal positions 
of all atoms have been optimized 
via the total energy and atomic forces minimization, while the lattice
parameters were fixed at the experimental values. 
For the FLAPW calculations cutoffs for the basis functions, 16.0 Ry, and 
potential representation, 81.0 Ry, and expansion in terms 
of spherical harmonics with $\ell \le$ 8 inside the muffin-tin spheres 
were used. 
The muffin-tin radii were 2.6 a.u. for Ca, 1.7 for Al and 1.5 for O atoms.
Summations over the Brillouin zone were carried out 
using 10-19 special {\bf k} points in the irreducible wedge.

\subsection{Electronic band structure}

\begin{figure*}
%\centerline{
%\includegraphics[height=4.1cm]{c.eps}
%\includegraphics[height=4.1cm]{3ca.eps}
%\includegraphics[height=4.1cm]{12c7a.eps}
%\includegraphics[height=4.1cm]{ca.eps}
%\includegraphics[height=4.1cm]{c2a.eps}
%\includegraphics[height=4.1cm]{c6a.eps}
%\includegraphics[height=4.1cm]{a.eps}
%\includegraphics[height=4.1cm]{a-m.eps}
%\includegraphics[height=4.1cm]{a-o.eps}
%
\includegraphics[height=7.1cm]{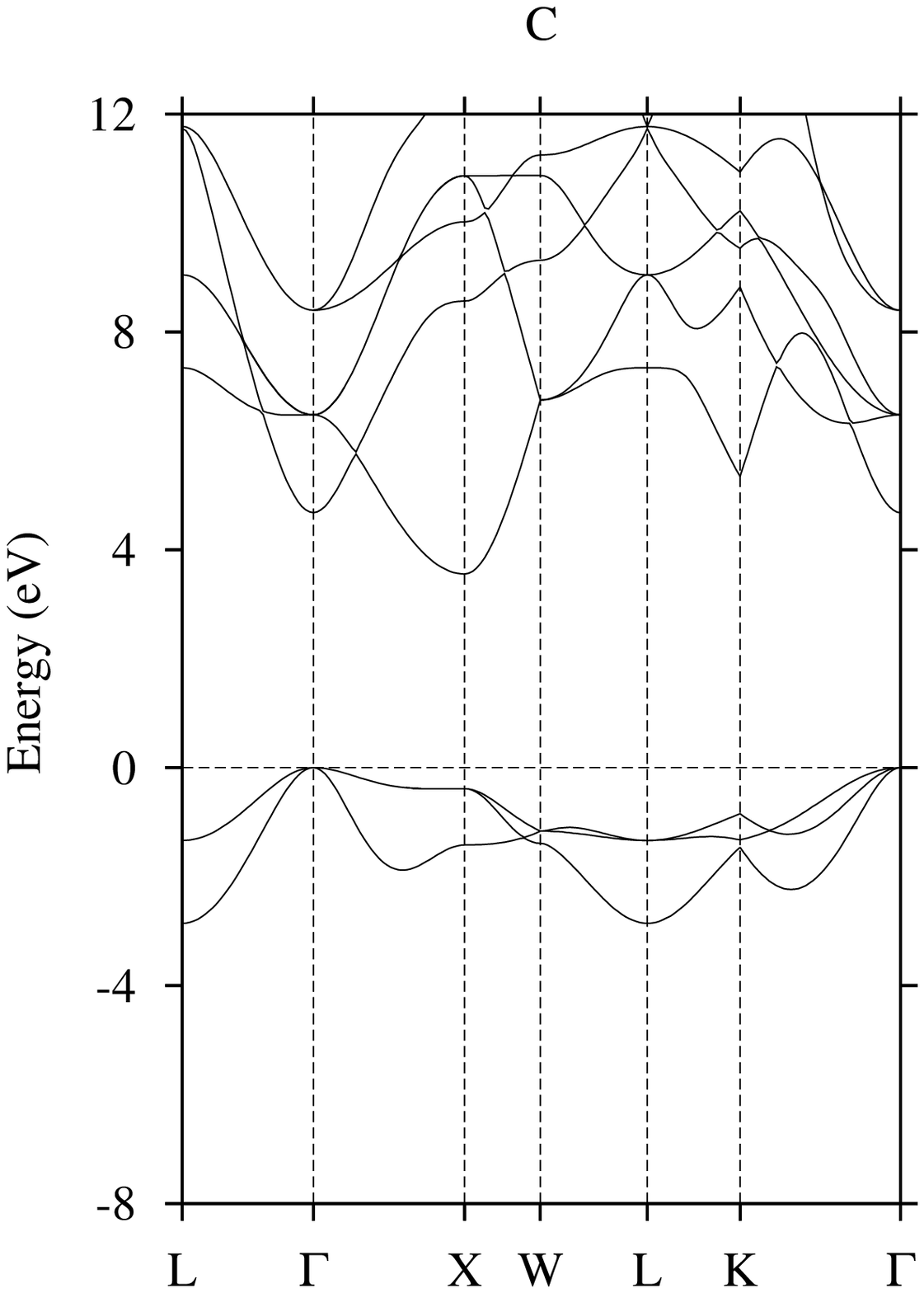}
\includegraphics[height=7.1cm]{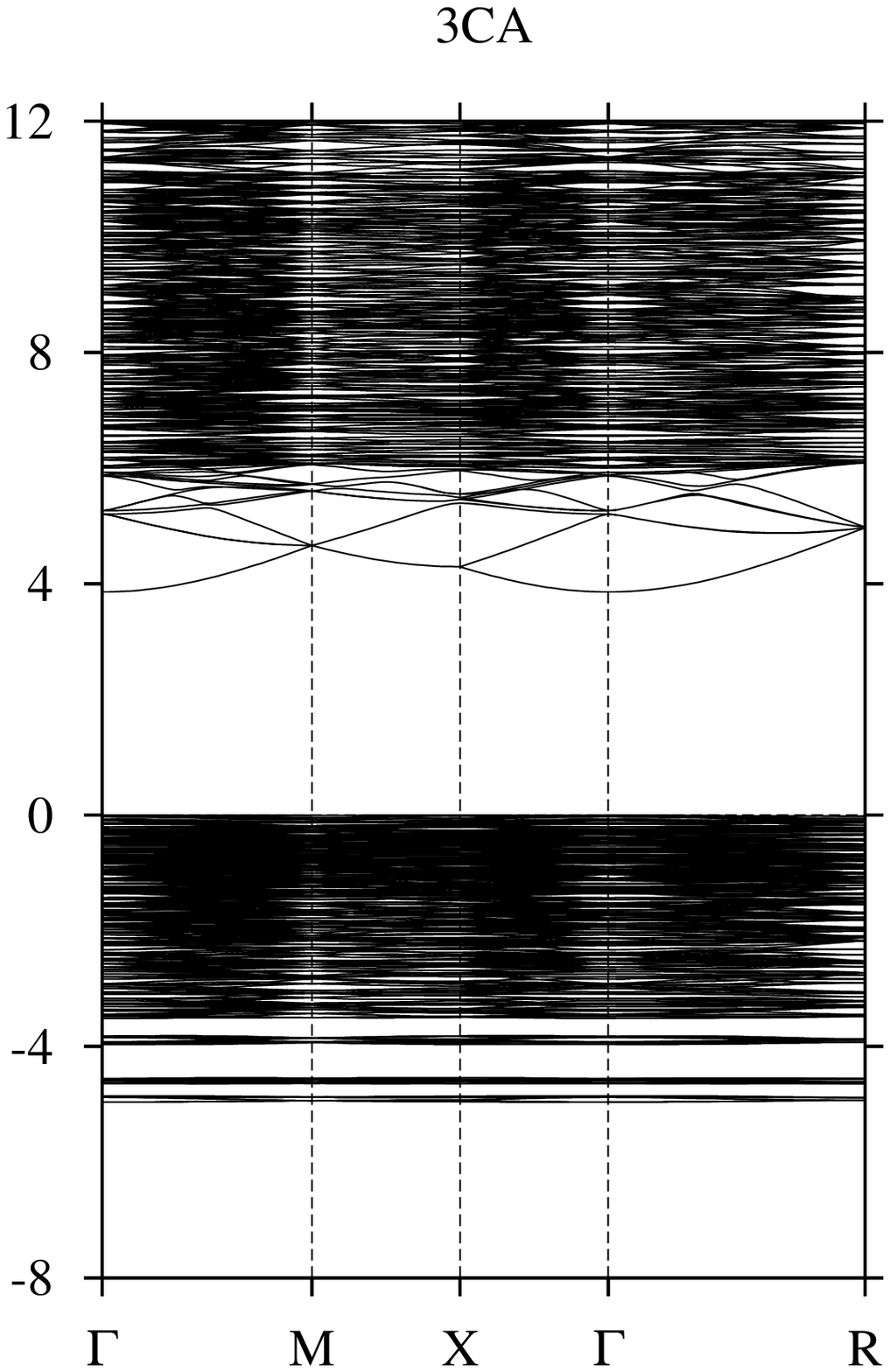}
\includegraphics[height=7.1cm]{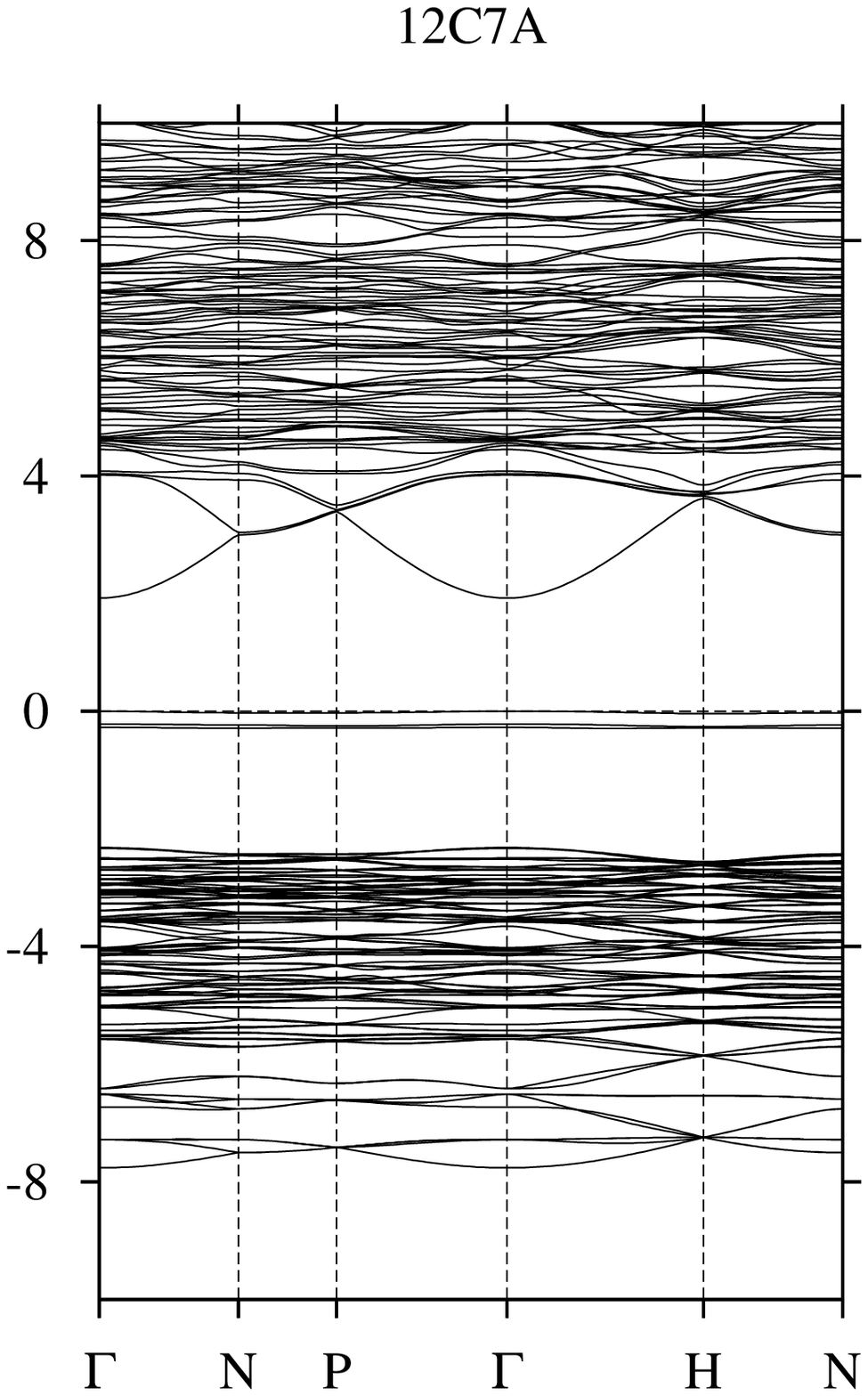}
\includegraphics[height=7.1cm]{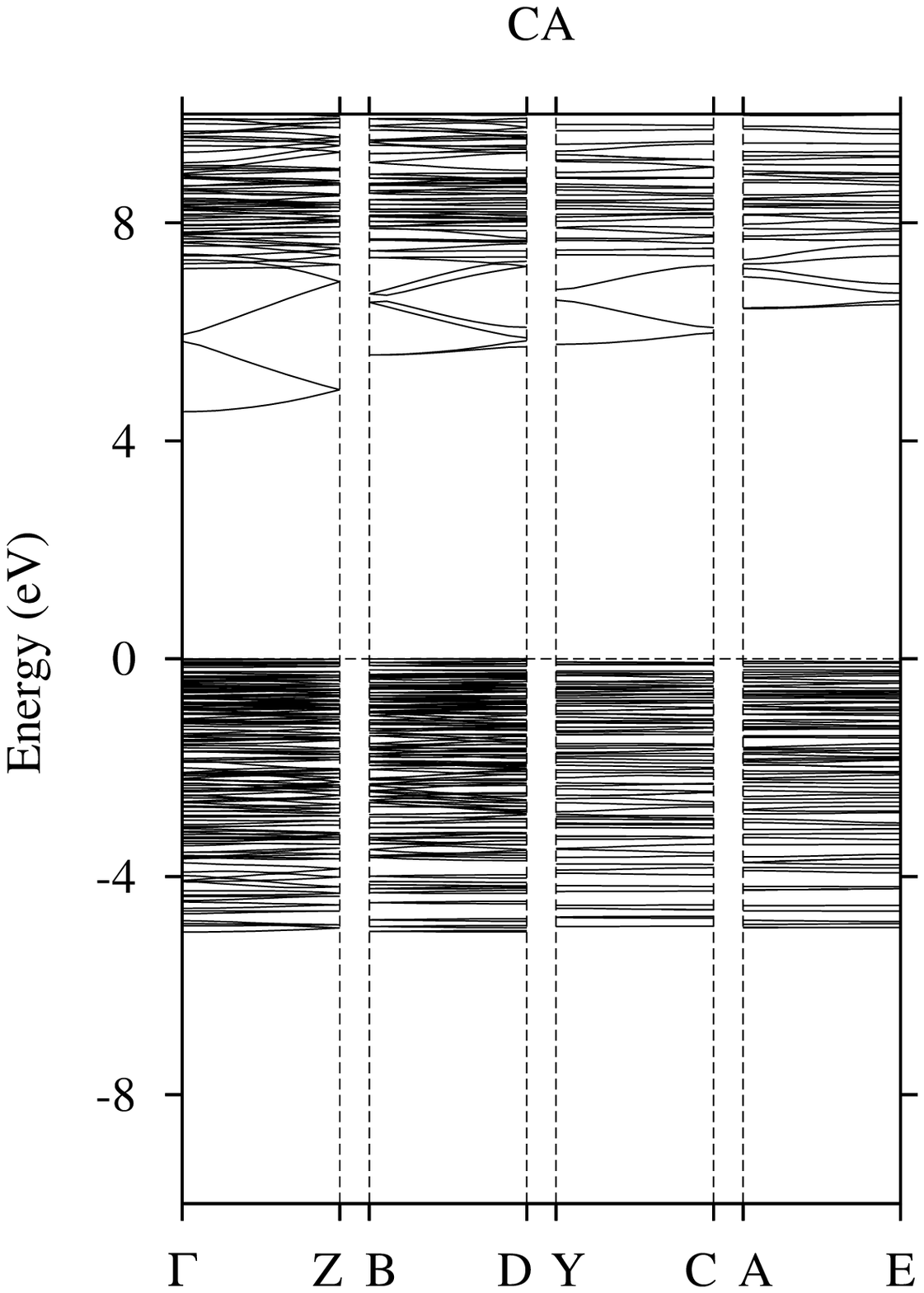}
\includegraphics[height=7.1cm]{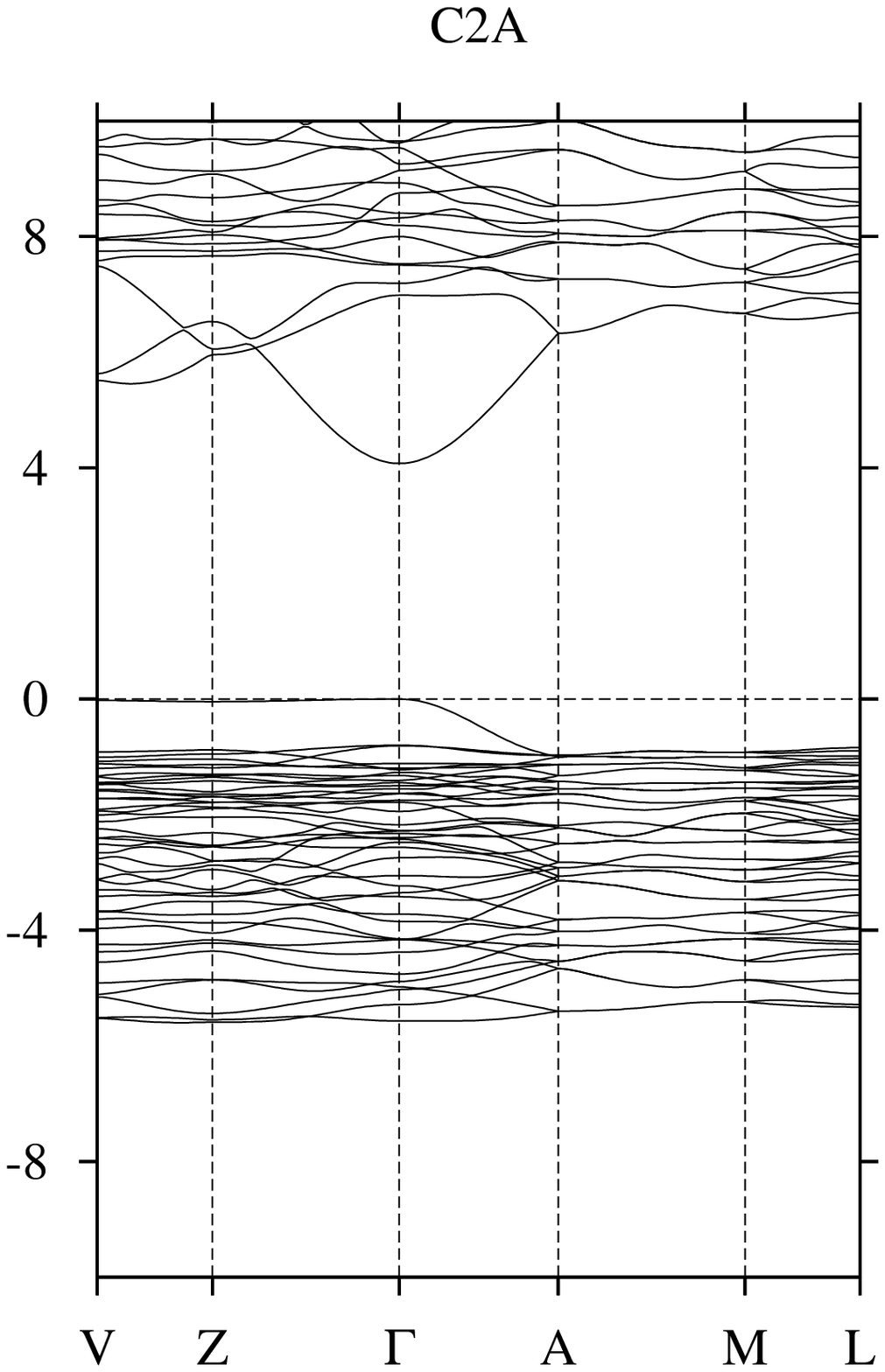}
\includegraphics[height=7.1cm]{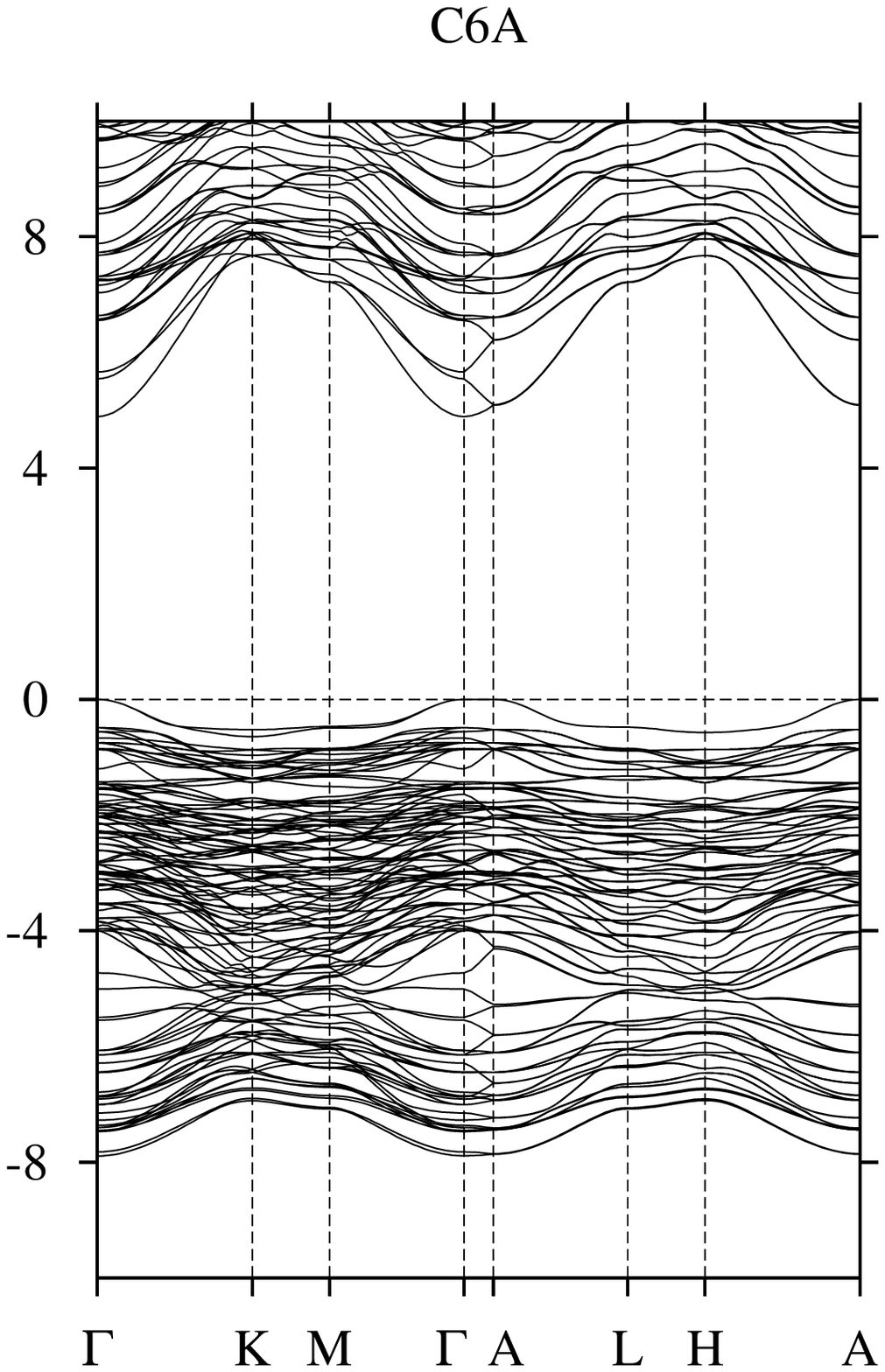}
\includegraphics[height=7.1cm]{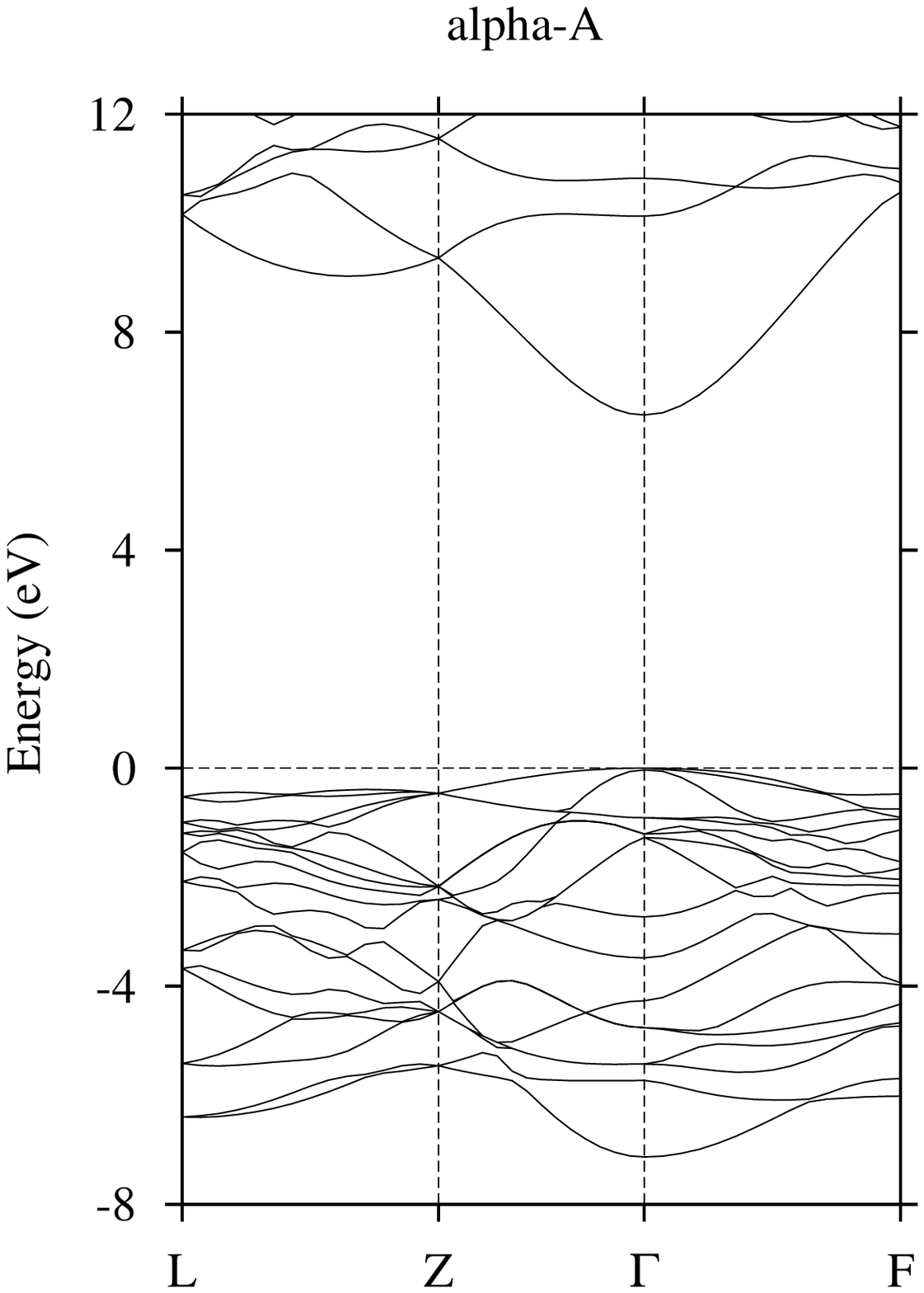}
\includegraphics[height=7.1cm]{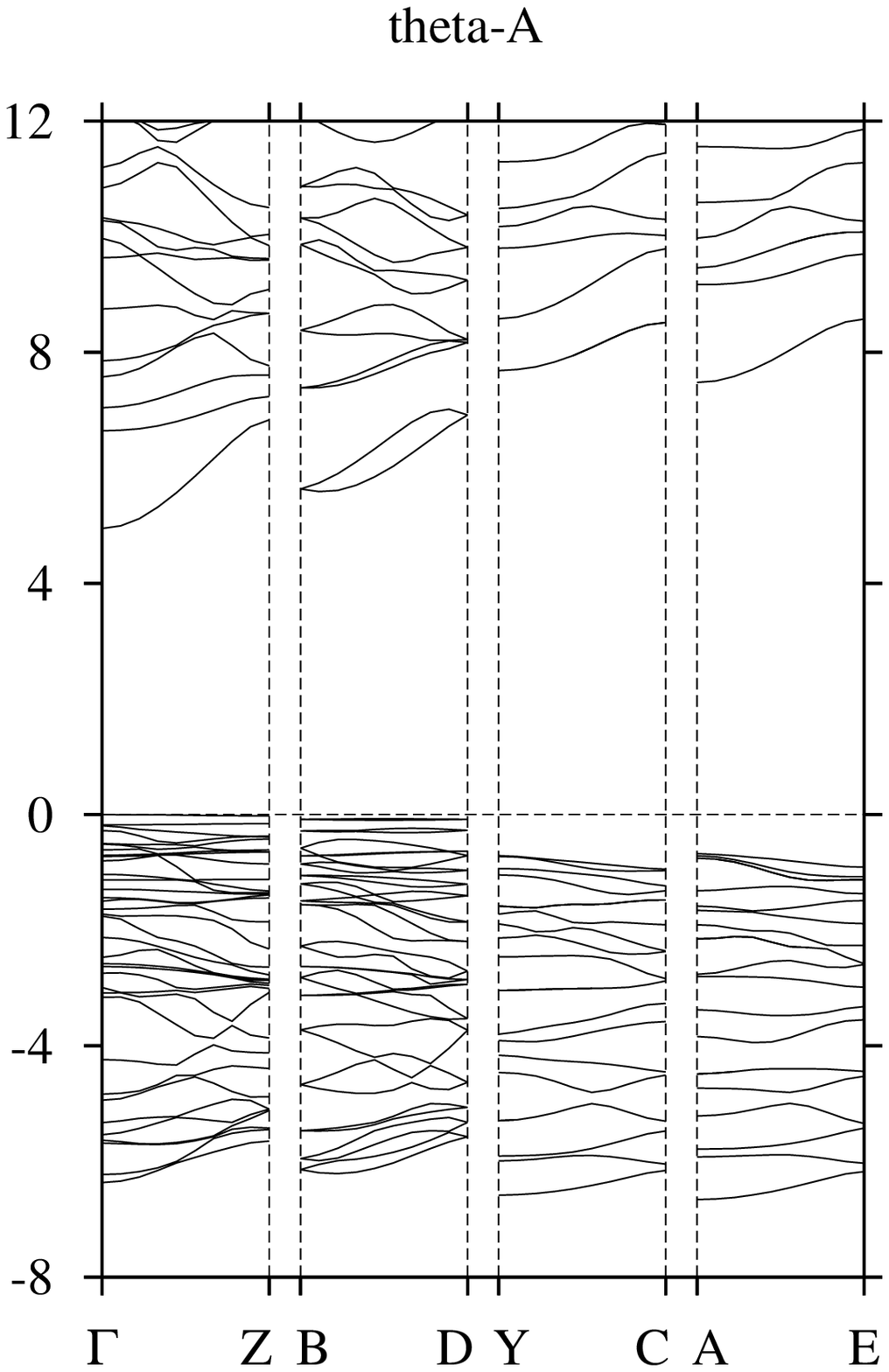}
\includegraphics[height=7.1cm]{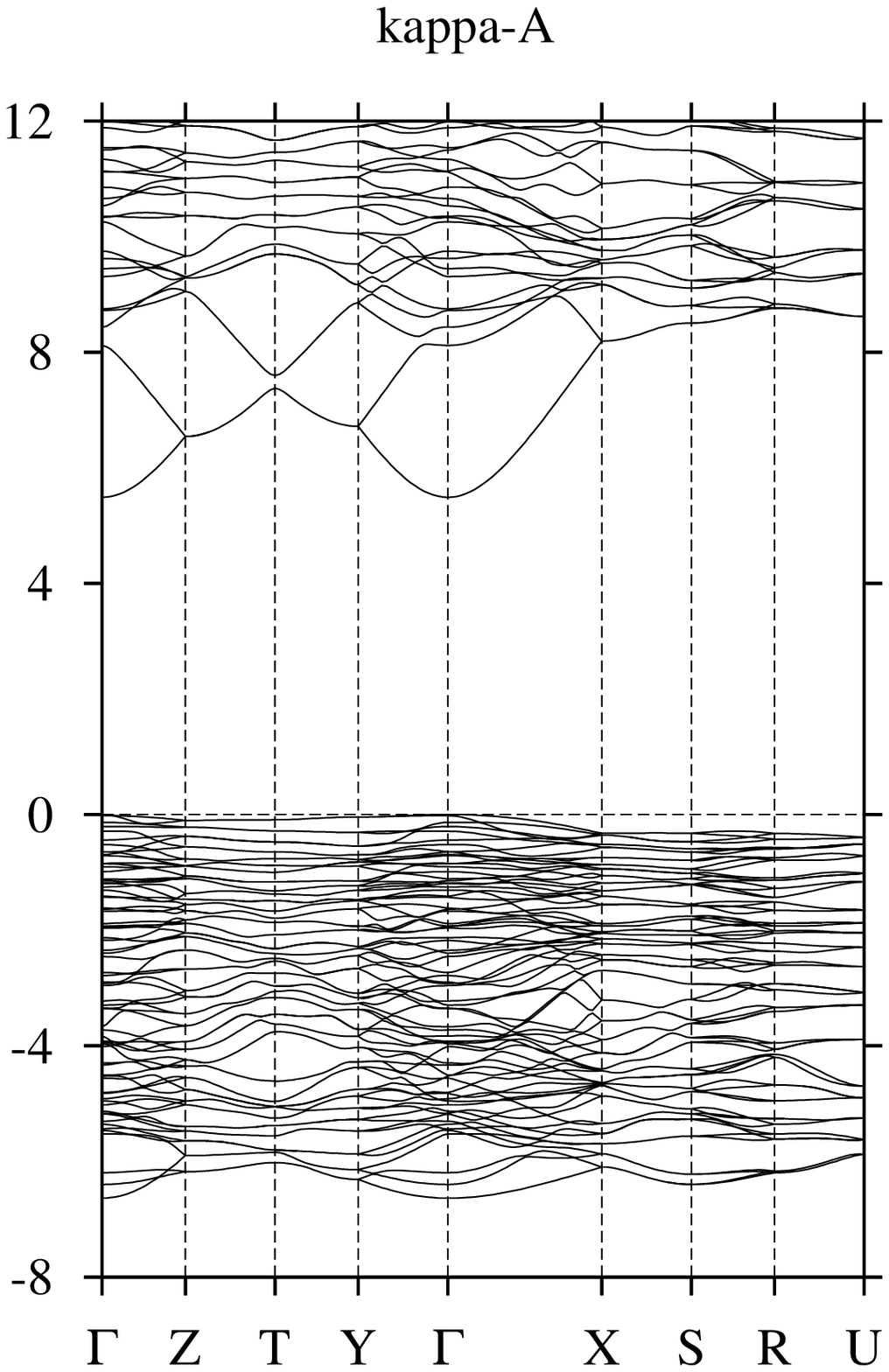}
%}
\caption{Electronic band structure plots for the compounds 
of the CaO-Al$_2$O$_3$ family and the terminal phases, CaO and $\alpha$-Al$_2$O$_3$.
Band structures of monoclinic $\theta$-Al$_2$O$_3$ and orthorhombic $\kappa$-Al$_2$O$_3$
are given for comparison.}
\label{bands}
\end{figure*}

The electronic band structures calculated along the high-symmetry 
directions in the corresponding Brillouin zones of the calcium aluminates 
are shown in Fig. \ref{bands}. All plots have the same energy 
scale so that the increase in the band gap value as the CaO 
(Al$_2$O$_3$) content decreases (increases), i.e., in the order 
C $<$ 3CA $<$ 12C7A $<$ CA $<$ C2A $<$ C6A $<$ $\alpha$-A, 
is clearly seen. Table \ref{table} lists the band gap 
and the valence band width (VBW) which also increases 
for the compounds with higher alumina content. 
The largest VBW value is found for the hexagonal C6A which has higher 
crystal symmetry
and hence provides the larger overlap between the 
orbitals of the neighboring atoms --
compared to those of the pure alumina phases (rhombohedral, 
monoclinic or orthorhombic, Table \ref{table}).

As expected, LDA underestimates the band gap in all oxides.
Our calculated band gap values are smaller 
by at least 1.45 eV for CaO (for the direct band gap at $\Gamma$ point), 
by 0.8 eV for C12A7 and by 2.3 eV for $\alpha$-Al$_2$O$_3$
as compared to the available experimental optical data.
The obtained band gaps are similar to the LDA results reported earlier
for CaO \cite{cao,cao1,cao2} and $\alpha$-Al$_2$O$_3$ \cite{al2o3}.
Note that the band gap underestimation does not affect the conclusions made.

It is widely accepted that the conductin band(s) in oxides of the main group 
metals is formed from the cations states, i.e., 
Ca $s$ and $d$ and Al $s$ and $p$ states.
However, our detailed analysis of the wavefunctions at the bottom 
of the conduction band, Table \ref{table-contrib}, 
provides a different picture: we find that 
the oxygen antibonding $p$-states give similar contributions
as compared to those from the cation(s) $s$-states \cite{p-contrib}. 
Further, we find that the relative atomic contributions are similar
within the room-temperature energy range, i.e., within $\sim$30 meV 
above the bottom of the conduction band.
Thus, both the cation(s) $s$ and anion $p$ states will be available for the transport of  
extra carriers and hence will determine the electron mobility 
in these materials once they are degenerately doped. 

It is important to point out that the density of states at the bottom of the conduction band 
is low due to the high dispersion $E(k)$ in this energy range 
which in turn originates in the large overlap between the wavefunctions 
of the neighboring 
cations and anions, i.e., the spherically symmetric $s$-orbitals and 
$p$-orbitals of 
the oxygen atoms. As discussed below, the low density of states will ensure 
the desired low optical absorption in degenerately doped materials.
Calcium $d$ and aluminum $p$ states give a significantly larger contributions 
to the density of states; however, because these states are located at much higher energies,
they will not be available for charge transport.

Another counter-intuitive finding is that the Ca and Al states give comparable contributions
to the bottom of the conduction band in these binary compounds.
This may come as a surprise because the band gap in Al$_2$O$_3$ 
is almost twice 
larger than the one in CaO, and so the unoccupied Al $s$-states 
are expected to be located deep in the conduction band.
Based on the analysis of the wavefunction at the bottom of the conduction band 
in several CaO-Al$_2$O$_3$ compounds calculated within highly precise FLAPW method, 
we find that the contributions from Al atoms  
are not negligible in 12C7A, C2A and C6A, Table \ref{table-contrib}.
Therefore, the states of {\it both} cations -- 
as well as the oxygen antibonding states -- 
will be available for extra electrons in properly doped materials.
Such a hybrid nature of the conduction band
may lead to a three-dimensional (isotropic) network for 
the electron transport in these complex multi-cation oxides -- 
that is consistent with the isotropic electron effective 
masses reported below.

\subsection{Electron effective mass}

Since the states of Ca and Al give comparable contributions 
to the conduction states in calcium aluminate oxides, 
both should contribute to the electron effective mass. 
Hence, for each binary compound one may expect 
an ``effective'' average over the effective masses of lime and alumina 
\cite{myEPL2}.
This may appear to be similar to the linear interpolation of the 
band gap and the electron effective mass within 
the virtual crystal approximation (VCA)
that has been known and widely utilized for semiconductor alloys, 
such as Ga$_{1-x}$In$_x$As or Ga$_{1-x}$Al$_x$N.
For alloys, however, the crystal lattice remains the same 
as the concentration of constituents is varied via substitution. 
In contrast, in calcium aluminates, the band gap and the effective 
mass averaging (which follows from VCA) is not justified because the lattice structure in these compounds
is dramatically different from the structure of the terminal phases.
Moreover, in all binary compounds with the exception for C6A, 
Al atoms have four oxygen neighbors while there is no alumina compound 
where all cations are 4-coordinated, Table \ref{table}.
Nevertheless, the increase in both the band gap and the electron effective mass 
as the content of Al$_2$O$_3$ increases is clearly seen from Table \ref{table}.

We point out here that the electron effective mass is directly related 
to the band gap value according to the {\bf k$\cdot$p} theory, namely,
the smaller the band gap, the smaller the electron effective mass and
vice versa \cite{underestim}.  However, the electron effective mass 
also depends on the overlap between the wavefunctions of the neighboring 
atoms, i.e., between the cation $s$-orbitals 
and the antibonding $p$-orbitals of the oxygen atoms.
Therefore, in addition to the oxygen coordination, 
the distortions in the polyhedra and in the cation-anion chains 
affect the orbital overlap and, hence, the electron effective mass.
This explains the fact that the effective mass is lower 
in orthorhombic $\kappa$-Al$_2$O$_3$ 
than in monoclinic alumina phase despite the opposite trend 
in the band gaps, 
Table \ref{table}. These two alumina phases can be compared to C6A where both 
four- and six-coordinated Al atoms are also present. In the latter, the electron effective mass
is the lowest due to the highest symmetry of its crystal structure
which provides the largest orbital overlap.

Note that CaO has indirect band gap with the conduction band minimum 
at $X$ point and the valence band maximum at $\Gamma$ point. 
The electron effective masses given in Table \ref{table} are calculated 
in the [100], [010] and [001] directions at $\Gamma$ point. For the directions 
in the standard Brillouin zone, i.e., [111] or $\Gamma$$L$ and [011] or $\Gamma$$K$, 
the electron effective masses are the same, 0.33 m$_e$. Thus, the effective mass is isotropic 
at $\Gamma$ point since the conduction band is parabolic at the wavevector $\vec{k}$=0, 
as expected. 
However, because $\Gamma$ point is $\sim$1 eV higher in energy 
with respect to 
the bottom of the conduction band, the mobility of extra electrons will be determined
by their effective mass at the $X$ point which we find to be 1.22 m$_e$ along
the $\Gamma$$X$ direction.

Significantly, the electron effective mass remains nearly isotropic 
in all CaO-Al$_2$O$_3$ compounds, Table \ref{table}, 
despite the structural complexity in these materials, namely, the low symmetry 
and thus local distortions in the cation-anion polyhedra; 
structural anisotropy due to irregular atomic arrangements such as
layers, rings, or chains of one type of cations; or the presence of large 
structural voids.
The largest deviation of the in-plane effective mass from 
the one calculated along 
the $z$ direction is found in monoclinic $\theta$-Al$_2$O$_3$
where the anisotropy factor
$\delta = (m_e^{[100]}+m_e^{[010]})/2m_e^{[001]}-1$ is 0.11. 
This finding is important from technological point of view. 
The isotropy of the electronic properties, i.e., the insensitivity of 
the electron effective mass to the oxygen coordination and structural 
variations, suggests that similar electronic properties can be achieved 
in a structure where the stoichiometry is maintained 
and the cations and anions alternate, i.e., 
cations are coordinated with anions and vice versa. 
Therefore, amorphous phases of calcium aluminates 
(which are readily available via the liquid solution, or sol-gel, 
preparation techniques \cite{Goktas,Gulgun,Tas})
can be successfully utilized. This is in marked contrast to 
amorphous Si where the directional interactions between 
the conduction $p$-orbitals give rise to a strong anisotropy 
of the transport properties 
and a significant decrease in the conductivity \cite{Hosono-Si}.

Finally, the conduction band topology in calcium aluminates resembles the one 
of the conventional \cite{my1epl} transparent conducting oxide hosts 
\cite{myreview,Woodward,Freeman,Mryasov,Asahi,CdO,myPRL2}: 
the high energy dispersion at the bottom of the conduction band indicates 
a small 
electron effective mass and hence should lead to a high carrier mobility 
upon degenerate doping of the materials. 
The electron effective masses found for the CaO-Al$_2$O$_3$ compounds 
are comparable to those in the well-known and commercially utilized 
transparent conductor hosts such as 
In$_2$O$_3$ (0.17 m$_e$ as obtained from an additional calculation 
within LMTO method for direct comparison) 
and ZnO (0.21 m$_e$ along the [100] and [010] directions and 0.19 m$_e$ 
along the [011] direction for the hexagonal phase, also calculated within LMTO).
Note that the mobility of extra carriers should play a crucial role 
in providing 
good electrical conductivity because large carrier concentrations 
(which may be 
challenging to achieve in the calcium aluminate oxides 
\cite{Neumark,VandeWalle,Zunger}) should be avoided to keep 
optical absorption low. The absorption arises 
due to the transitions from the conduction band partially occupied 
by introduced electrons and due to the plasma frequency of the free carriers.
In addition, currently known conventional transparent conductors (oxides of 
post-transition metals such as In, Zn, Sn, Ga and Cd) 
possess a relatively small band gaps of $\sim$2-3 eV.
Therefore, large carrier concentrations may be required 
to provide the desired optical transmittance in the short-wavelength 
range of the visible part of the spectrum
that is attained via a pronounced displacement of the Fermi energy 
(so-called Burstein-Moss shift). 
This may not be required in calcium aluminates where significantly 
larger band gaps will ensure that 
the intense transitions from the valence band are out of the visible range.

\subsection{Hole effective mass}

The electronic band structure plots, Fig. \ref{bands}, reveal another interesting peculiarity, 
namely, the large anisotropy of the top of the valence band 
and, hence, of the hole effective mass, Table \ref{table}. 
The anisotropy factor, calculated as 
$\delta = (m_h^{[100]}+m_h^{[010]})/2m_h^{[001]}$, 
is equal to 10.71, 0.05 and 11.83 for C2A and for 
monoclinic ($\theta$) and orthorhombic ($\kappa$) 
Al$_2$O$_3$, respectively. The large anisotropy is also observed 
for CA and C6A, however, we could not determine the hole effective masses 
in all three directions due to the non-dispersive
character of the bands along some of the directions in these compounds, 
Table \ref{table}.
Small values of the hole effective mass obtained for C2A and C6A, 
i.e., 0.6-0.7 m$_e$, should stimulate the search for efficient ways to convert 
these oxides into highly desirable p-type conductors with mobile carriers -- 
a complementary characteristic to the n-type behavior discussed above.

\subsection{Conclusions}

Based on the electronic band structure investigations of
calcium aluminates with different composition and structure,
we find that anions and both cations give comparable
contributions to the conduction states 
that may lead to isotropic three-dimensional distribution of 
the conduction electron density if these materials are properly doped. 
It is important to note, however, that doping of a structurally anisotropic
material (such as 3CA, C6A or C2A) may result in a non-uniform distribution
of carrier donors such as oxygen defects or aliovalent substitutional dopants.
Therefore, whether or not the isotropic behavior of the host material 
is maintained will depend on the carrier generation mechanism.
Amorphous counterparts of the cement phases readily offer a way 
to attain isotropic transport. In addition, 
due to the low electron effective mass which is shown to be 
insensitive to structural distortions and disorder, extra carriers in the amorphous oxides
are expected to be nearly as mobile as they are in the crystalline phases, 
making the materials attractive from the technological point of view.

%\newpage

\begin{table*}
\begin{tabular}{l|c|c|cc|cc|c|c|ccc|ccc} \hline
 & Space group & Z & $\langle$D$_{\rm Ca-O}\rangle$ & Ca n.n. & $\langle$D$_{\rm Al-O}\rangle$ & Al n.n. & E$_g$ & VBW & m$_e^{[100]}$ & m$_e^{[010]}$ & m$_e^{[001]}$ & m$_h^{[100]}$ & m$_h^{[010]}$ & m$_h^{[001]}$ \\ \hline
C     & $Fm\bar{3}m$ &  1 & 2.40 &  6  &  --- & ---                  &  3.55  & 2.8 & 0.33 & 0.33 & 0.33 & 1.25 & 1.25 & 1.25             \\
3CA   & $Pa\bar{3}$  & 24 & 2.42 &  6  & 1.75 & 4                    &  3.86  & 3.5 & 0.33 & 0.33 & 0.33 & $\infty$ & $\infty$ & $\infty$ \\
12C7A & $I\bar{4}3d$ &  1 & 2.43 &  6  & 1.75 & 4                    &  4.23  & 5.4 & 0.37 & 0.37 & 0.37 & 2.54 & 2.54 & 2.54             \\
CA    & $P2_1/n$     & 12 & 2.51 &  6  & 1.75 & 4                    &  4.53  & 5.0 & 0.39 & 0.38 & 0.37 & $\infty$ & 3.10 & $\infty$     \\
C2A   & $C2/c$       &  2 & 2.39 &  5  & 1.76 & 4                    &  4.87  & 5.6 & 0.38 & 0.38 & 0.40 & 6.64 & 6.64 & 0.62             \\
C6A   & $P6_3/mmc$   &  2 & 2.71 &  6  & 1.80, 1.93, 1.91 & 4, 5, 6  &  5.38  & 7.8 & 0.34 & 0.34 & 0.31 & 0.68 & 0.68 & $\infty$         \\
$\alpha$-A     
      & $R\bar{3}c$  &  2 &  --- & --- & 1.91 &  6                   &  6.48  & 7.1 & 0.38 & 0.38 & 0.38 & 3.99 & 3.99 & 3.99             \\ 
$\theta$-A  
      & $C2/m$       &  2 &  --- & --- & 1.77, 1.93 & 4, 6           &  4.95  & 6.4 & 0.41 & 0.41 & 0.37 & 0.64 & 0.64 & 13.68             \\ 
$\kappa$-A  
      & $Pna2_1$     &  8 &  --- & --- & 1.77, 1.94 & 4, 6           &  5.49  & 6.6 & 0.37 & 0.35 & 0.36 & 4.90 & 6.22 & 0.47             \\ \hline
\end{tabular}
\caption{Structural and electronic properties of the 
CaO-Al$_2$O$_3$ compounds.
Crystal space group; the number of formula units per unit cell, Z, 
used in the calculations; the average cation-anion distances,
$\langle$D$_{\rm Ca-O}\rangle$ and $\langle$D$_{\rm Al-O}\rangle$, in \AA;
the number of nearest oxygen neighbors for Ca and Al atoms, n.n.;
band gap, E$_g$, in eV; valence band width, VBW, in eV;
and the electron and hole effective masses, m$_e$ and m$_h$, in units of the electron mass, 
calculated along the specified crystallographic directions.
%and the averaged effective mass, $\langle$m$\rangle$, estimated from Eq. (1) using the masses 
%of the corresponding single metal oxides.
}
\label{table}
\end{table*}

%\begin{equation}
%m*(xCaO \cdot yAl_2O_3) = x \cdot m*(CaO) + y \cdot m*(Al_2O_3)
%\end{equation}

\begin{table}
\begin{tabular}{lccc} \hline
Compound & C$^{\Gamma}_{Ca}$ & C$^{\Gamma}_{Al}$ & C$^{\Gamma}_{O}$ \\ \hline
CaO                       & 59 & -- & 41 \\
12CaO$\cdot$7Al$_2$O$_3$  & 64 &  7 & 29 \\
CaO$\cdot$2Al$_2$O$_3$    & 49 & 16 & 34 \\
CaO$\cdot$6Al$_2$O$_3$    & 42 & 25 & 33 \\
$\alpha$-Al$_2$O$_3$      & -- & 55 & 45 \\ \hline
\end{tabular}
\caption{Relative average atomic contributions
to the conduction band wavefunction at $\Gamma$ point, in per cent,
as calculated for several oxides within FLAPW method.
}
\label{table-contrib}
\end{table}


\begin{thebibliography}{99}
\bibitem{Hosono-original}
K. Hayashi, S. Matsuishi, T. Kamiya, M. Hirano, H. Hosono,
%Light-induced conversion of an insulating refractory oxide
%into a persistent conductor.
Nature {\bf 419}, 462
%-465
(2002).
\bibitem{Bertoni-thesis}
M. Bertoni, 
Ph.D. Thesis, Northwestern University, Evanston, IL, USA, 2006. 
\bibitem{Kim}
S.W. Kim, S. Matsuishi, T. Nomura, Y. Kubota, M. Takata, K. Hayashi, T. Kamiya, M. Hirano, and H. Hosono,
%Metallic State in a Lime-Alumina Compound with Nanoporous Structure
Nano Lett. {\bf 7}, 1138
%-1143
(2007).
\bibitem{HosonoAdvanced}
Y. Toda, S. Matsuishi, K. Hayashi, K. Ueda, T. Kamiya, M. Hirano, and H. Hosono,
%Field Emission of Electron Anions Clathrated in Subnanometer-Sized Cages in [Ca24Al28O64]4+(4e-)
Adv. Mater. {\bf 16}, 685
%-689 
(2004).
\bibitem{Science}
S. Matsuishi, Y. Toda, M. Miyakawa, K. Hayashi, T. Kamiya,
M. Hirano, I. Tanaka, and H. Hosono,
Science {\bf 301}, 626 (2003).
\bibitem{myPRL}
J.E. Medvedeva, A.J. Freeman, M.I. Bertoni, T.O. Mason,
Phys. Rev. Lett. {\bf 93}, 16408 (2004)
\bibitem{Kim123}
S.W. Kim, M. Miyakawa, K. Hayashi, T. Sakai, M. Hirano, and H. Hosono, 
%Simple and efficient fabrication of room temperature stable electride: 
%Melt-solidification and glass ceramics
J. Amer. Chem. Soc. {\bf 127}, 1370
%1371 
(2005); 
%
S.W. Kim, Y. Toda, K. Hayashi, M. Hirano, and H. Hosono,
%Synthesis of a room temperature stable 12CaO center dot 7Al(2)O(3) electride 
%from the melt and its application as an electron field emitter
Chem. Mater. {\bf 18}, 1938
%-1944 
(2006); 
%
S.W. Kim, K. Hayashi, M. Hirano, H. Hosono, and I. Tanaka,
%Electron carrier generation in a refractory oxide 12CaO (center dot) 7Al(2)O(3) 
%by heating in reducing atmosphere: Conversion from an insulator to a persistent conductor
J. Amer. Ceram. Soc. {\bf 89}, 3294
%-3298 
(2006).
\bibitem{myAPL}
J.E. Medvedeva, A.J. Freeman,
Appl. Phys. Lett. {\bf 85}, 955 (2004).
\bibitem{MarianaJAP}
M.I. Bertoni, T.O. Mason, J.E. Medvedeva, A.J. Freeman, K.R. Poeppelmeier, B. Delley,
J. Appl. Phys. {\bf 97}, 103713 (2005).
\bibitem{MarianaSi}
M. Bertoni, J.E. Medvedeva, Y.Q. Wang, A. Freeman, K.R. Poeppelmeier, and T.O. Mason,
%Enhanced electronic conductivity in Si-substituted calcium aluminate
J. Appl. Phys.
\bibitem{my1epl}
J.E. Medvedeva, A.J. Freeman,
%Combining high conductivity with complete optical transparency: A band-structure approach.
Europhys. Lett. {\bf 69}, 583
%-587
(2005).
\bibitem{myreview}
J.E. Medvedeva, 
%Special issue on Transparent Conducting Oxides, 
%guest editors G. Kiriakidis and S. Mao,
Appl. Phys. A (2007).
\bibitem{Rankin}
E.S. Shepherd, G.A. Rankin, and F.E. Wright, 
%The Binary Systems of Alumina with Silica, Lime and Magnesia, 
Am. J. Sci. {\bf 28}, 293
%-333 
(1909);
%
G.A. Rankin, and F.E. Wright, 
%The Ternary System CaO-Al2O3-SiO2, 
Am. J. Sci., {\bf 39}, 1
%-79 
(1915).
\bibitem{Wisnyj}
L.G. Wisnyi, {\it The High Alumina Phases in the System Lime-Alumina}, 
Ph.D. Thesis, Rutgers University, New Brunswick, NJ, USA, 1955. 
\bibitem{Barnes}
{\it Structure and performance of cements}, edited by P. Barnes 
(Applied Science Publishers, 1983).
\bibitem{Morozova}
L.P. Morozova, F.D. Tamas, and T.V. Kuznetsova, 
%Preparation of Calcium Aluminates by a Chemical Method, 
Cement and Concrete Research {\bf 18}, 375
%-388 
(1988). 
\bibitem{Hallstedt}
B. Hallstedt, 
%Assessment of the CaO-Al2O3 System, 
J. Am. Cer.  Soc. {\bf 73}, 15
%-23 
(1990).
\bibitem{Singh}
V.K. Singh, M.M. Ali, and U.K. Mandal,
%FORMATION KINETICS OF CALCIUM ALUMINATES 
J. Amer. Ceram. Soc. {\bf 73}, 872
%-876
(1990). 
\bibitem{Goktas}
A.A. Goktas, and M.C. Weinberg, 
%Preparation and  Crystallization of  Sol-Gel Calcia-Alumina Compositions, 
J. Am. Ceram. Soc. {\bf 74}, 1066
%-1070 
(1991).
\bibitem{Eriksson}
G. Eriksson, and A.D. Pelton, 
%Critical Evaluation and  Optimization of the Thermodynamic Properties and Phase Diagrams  
%of the CaO-Al2O3, Al2O3-SiO2, and CaO-Al2O3-SiO2 Systems, 
Metall. Trans. B {\bf 24B}, 807
%-816 
(1993). 
\bibitem{Gulgun}
M.A. Gulgun, O.O. Popoola, W.M. Kriven, 
%Chemical Synthesis and Characterization of Calcium Aluminate Powders, 
J. Am. Ceram. Soc. {\bf 77}, 531
%-539 
(1994).
\bibitem{Tas}
A.C. Tas. 
%Chemical Preparation of the Binary Compounds of CaO-Al2O3 System 
%by Self-propagating Combustion Synthesis, 
J. Amer. Ceram. Soc. {\bf 81}, 2853
%-2863 
(1998).
\bibitem{Yi}
H.C. Yi, J.Y. Guign\'{e}, J.J. Moore, F.D. Schowengerdt, L.A. Robinson, and A.R. Manerbino,
J. Mater. Science {\bf 37}, 4537
%-4543(7)
(2002).
%Preparation of calcium aluminate matrix composites by combustion synthesis
\bibitem{Kopanda}
J.E. Kopanda, and G. Mac Zura, 
%Production Processes, Properties, and  Applications for Calcium Aluminate  Cements; 
%pp. 171-184 
{\it Alumina Chemicals Science and Technology Handbook}, edited by  L.D. Hart 
(American Ceramic Society, Westerville, OH, 1990), p. 171.
\bibitem{Shelby}
%Calcium fluoroaluminate glasses
J.E. Shelby, C.M. Shaw, and M.S. Spess,
J. Appl. Phys. {\bf 66}, 1149
%-1154 
(1989).
\bibitem{Jeev}
J. Jeevaratnam, F.P. Glasser, and L.S. Dent Glasser, 
%Anion Substitution and Structure of 12CaO.7Al2O3, 
J. Am. Ceram. Soc. {\bf 47} 105
%-106 
(1964).
\bibitem{Chatterji}
A.K. Chatterjee, and G.I. Zhmoidin, 
%The Phase Equilibrium Diagram  of the System CaO-Al2O3-CaF2, 
J. Mater. Sci. {\bf 7}, 93
%-97 
(1972). 
\bibitem{Williams}
P.P. Williams, 
%Refinement of the Structure of 11CaO.7Al2O3.CaF2, 
Acta Cryst. {\bf B29}, 1550
%-1551 
(1973).
\bibitem{Zhmoidin}
G.I. Zhmoidin, and G.S. Smirnov. 
%Characteristics of the crystals of derivatives of 12CaO·7Al2O3, 
Inorganic Mater. {\bf 18}, 1595
%-1601
(1982).
\bibitem{maySi}
%Oxidative destruction of hydrocarbons on a new zeolite-like crystal 
%of Ca12Al10Si4O35 including O-2(-) and O-2(2-) radicals
S. Fujita, K. Suzuki, M. Ohkawa, T. Mori, Y. Iida, Y. Miwa, H. Masuda, and S. Shimada,
Chem. Mater. {\bf 15}, 255
%-263 JAN 14 
(2003);
%
%Controlling the quantity of radical oxygen occluded in a new aluminum silicate with nanopores
S. Fujita, M. Ohkawa, K. Suzuki, H. Nakano, T. Mori, and H. Masuda,
Chem. Mater. {\bf 15} 4879
%-4881 DEC 30 
(2003). 
\bibitem{Mondal}
P. Mondal, and J.W. Jeffery. 
%The crystal structure of tricalcium aluminate, Ca3Al2O6, 
Acta Cryst. {\bf B31}, 689
%-697
(1975).
\bibitem{Utsunomiya}
A. Utsunomiya, K. Tanaka, H. Morikawa, F. Marumo, and H. Kojima, 
%Structure refinement of CaO·6Al2O3, 
J. Solid State Chem. {\bf 75}, 197
%-200 
(1988).
\bibitem{Du}
L.-S. Du, and J.F. Stebbins, 
%Calcium and Strontium Hexaluminates: NMR Evidence that "Pentacoordinate" Cation 
%Sites Are Four-Coordinated, 
J. Phys. Chem. {\bf 108}, 3681
%-3685
(2004).
\bibitem{Hofmeister}
A.M. Hofmaister, B. Wopenka, and A.J. Locock,
%Spectroscopy and structure of hibonite, grossite and CaAl2O4...
Geohimica et Cosmochimica Acta {\bf 68}, 4485
%-4503
(2004).
\bibitem{LMTO}
O. K. Andersen, O. Jepsen, M. Sob,
{\it Electronic Band Structure and its Applications,}
(ed. M. Yussouff. Springer-Verlag, Berlin, 1986).
\bibitem{Christensen}
A.N. Christensen. 
%Neutron powder diffraction profile refinement studies on Ca11.3Al14O32.3 
%and CaClO(D0.88H0.12), 
Acta Chemica Scandinavica {\bf A41}, 110
%-112
(1987).
\bibitem{Horkner}
W. H\"{o}rkner, and H.K. M\"{u}ller-Buschbaum,
%Zur Krsit. CaAl2O4
J. Inorg. Nucl. Chem. {\bf 38}, 983
%-984
(1976).
\bibitem{Goodwin}
D.W. Goodwin, and A.J. Lindop,
%The crystal structure of CaO.2Al2O3
Acta Cryst. B {\bf 26}, 1230
%-1235
(1970).
\bibitem{Husson}
E. Husson E, and Y. Repelin,
%Structural studies of transition aluminas. Theta alumina
Euro. J. Solid State Inorg. Chem. {\bf 33} 1223
%-1231
(1996).
\bibitem{Ollivier}
B. Ollivier, R. Retoux, P. Lacorre, D. Massiot, and G. Ferey,
%Crystal structure of kappa-alumina: an X-ray powder diffraction,
% TEM and NMR study
J. Mater. Chem. {\bf 7} 1049
%-1056  Issue 6
(1997).

\bibitem{FLAPW}
E. Wimmer, H. Krakauer, M. Weinert, A.J. Freeman, 
%Full-potential self-consistent linearized-augmented-plane-wave method 
%for calculating the electronic structure of molecules and surfaces -- O$_2$ molecule.
Phys. Rev. B {\bf 24}, 864 (1981)
%-875 (1981).
\bibitem{FLAPW1}
M. Weinert, E. Wimmer, A.J. Freeman, 
%Total-energy all-electron density functional method 
%for bulk solids and surfaces.
Phys. Rev. B {\bf 26}, 4571 (1982)
%-4578 (1982).
\bibitem{cao}
S.K. Medeiros, E.L. Albuquerque, F.F. Maia Jr, J.S. de Sousa, E.W.S. Caetano, and V.N. Freire,
J. Phys. D: Appl. Phys. {\bf 40}, 1655
%-1658
(2007).
%CaO first-principles electronic properties and MOS device simulation
\bibitem{cao1}
A. Yamasaki, T. Fujiwara, Phys. Rev. B {\bf 66}, 245108 (2002).
\bibitem{cao2}
H. Baltache, R. Khenata, M. Sahnoun, M. Driz, B. Abbar, and B. Bouhafs, 
% Full potential calculation of structural, electronic and elastic properties 
%of alkaline earth oxides MgO, CaO and SrO 
Physica B {\bf 344}, 334 (2004). 
%-342
\bibitem{al2o3}
A.A. Demkov, L.R.C. Fonseca, E. Verret, J. Tomfohr, and O.F. Sankey,
% Complex band structure and the band alignment problem at the Si–high-k dielectric interface
Phys. Rev. B {\bf 71}, 195306 (2005).
%-8
\bibitem{p-contrib}
The presence of the oxygen $p$-states in the conduction band of MgO and MgS 
has been pointed out by de Boer and de Groot \cite{Boer}. 
A detailed comparison with CaO and CaS was later done 
by Basalaev {\it et al} \cite{Basalaev} who found that the energy location of the 
unoccupied $d$ and $p$ states of cations affects the relative anion-cation contributions.
The role of the antibonding oxygen states in the conduction states of 
a variety of main group oxides
has been also stressed by Mizoguchi and Woodward \cite{Woodward}.
\bibitem{Boer}
P.K. de Boer, R.A. de Groot, J. Phys. Condens. Matter {\bf 10}, 10241-10248 (1998).
%The conduction bands of MgO, MgS and HfO$_2$.
\bibitem{Basalaev}
Y.M. Basalaev, Y.N. Zhuravlev, A.V. Kosobutskii, and A.S. Poplavnoi, 
%The genesis of energy bands formed by sublattice states in
%alkaline-earth metal oxides and sulfides
Phys. Solid State {\bf 46}, 848 (2004).
%-852
\bibitem{Woodward}
H. Mizoguchi, and P.M. Woodward,
%Electronic structure studies of main group oxides possessing edge-sharing octahedra: 
%Implications for the design of transparent conducting oxides. 
Chem. Mater. {\bf 16}, 5233 (2004).
%-5248 (2004). 
%\bibitem{Dargam}
%T.G. Dargam, R.B. Capaz, B. Koiller
%Critical analysis of the virtual crystal approximation
%Brazilian J. Phys. {\bf 27A}, 299
%-304
%(1997).
%\bibitem{Sokeland}
%Density functional and quasiparticle band-structure calculations for GaxAl1-xN and GaxIn1-xN alloys
%F. S\"{o}keland, M. Rohlfing, P. Kr\"{u}ger, and J. Pollmann,
%Phys. Rev. B {\bf 68}, 075203 (2003).
%-11
\bibitem{myEPL2}
J.E. Medvedeva, 
%Averaging of the electron effective mass in multicomponent transparent conducting oxides.
Europhys. Lett. {\bf 78}, 57004
%-6
(2007).
\bibitem{underestim}
This also implies that because
LDA underestimates the band gap values, the electron effective 
masses are also underestimated.
\bibitem{Hosono-Si}
K. Nomura, H. Ohta, A. Takagi, T. Kamiya, M. Hirano, and H. Hosono,
%Room temperature fabrication of transparent flexible thin-film transistors
%using amorphous oxide semiconductors
Nature, {\bf 432}, 488 (2004).
%-492
\bibitem{Freeman}
A.J. Freeman, 
%{\it et al}, 
K.R. Poeppelmeier, T.O. Mason, R.P.H. Chang, and T.J. Marks, 
%Chemical and Thin-Film Strategies for New Transparent Conducting Oxides
MRS Bull. {\bf 25}, 45 (2000).
\bibitem{Mryasov}
O.N. Mryasov and A.J. Freeman, Phys. Rev. B {\bf 64}, 233111 (2001).
\bibitem{Asahi}
R. Asahi, 
%{\it et al},
A. Wang, J. R. Babcock, N. L. Edleman, A. W. Metz, M. A. Lane, 
V. P. Dravid, C. R. Kannewurf, A. J. Freeman, and T. J. Marks,
%First-principles calculations for understanding high conductivity 
%and optical transparency in InxCd1−xO films
Thin Solid Films {\bf 411}, 101 (2002).
%-105
\bibitem{CdO}
Y. Yang, 
%{\it et al},
S. Jin, J.E. Medvedeva, J.R. Ireland, A.W. Metz, J. Ni, 
M.C. Hersam, A.J. Freeman, and T.J. Marks,
%CdO as the Archetypical Transparent Conducting Oxide. Systematics of 
%Dopant Ionic Radius and Electronic Structure Effects on Charge Transport and Band Structure
J. Amer. Chem. Soc. {\bf 127}, 8796 (2005).
\bibitem{myPRL2}
J.E. Medvedeva, Phys. Rev. Lett. {\bf 97}, 086401 (2006).
\bibitem{Neumark}
G.F. Neumark,
%Defects in wide band gap II-VI crystals.
Mater. Sci. Eng. R {\bf 21}, 1 (1997).
%-46 (1997). 
\bibitem{VandeWalle}
C.G. Van de Walle,
%Strategies for controlling the conductivity of wide-band-gap semiconductors.
Phys. Stat. Solidi B {\bf 229}, 221 (2002).
%-228 (2002).
\bibitem{Zunger}
A. Zunger,
%Practical doping principles. 
Appl. Phys. Lett. {\bf 83}, 57 (2003).
%-59 (2003).
\end{thebibliography}
\end{document}